\documentclass[prd,twocolumn,nofootinbib,showpacs,eqsecnum,floatfix,superscriptaddress]{revtex4}
\usepackage{latexsym,amsbsy}
  \usepackage[dvips]{graphicx}
 \DeclareGraphicsExtensions{.eps, .jpg}

\begin{document}
\title{Contribution to hadrons' width from classical instability
of Y configuration\\ and other string hadron models}
\author{G. S. Sharov}
\email{german.sharov@mail.ru} \affiliation{Tver state university, 170002, Sadovyj per.
35, Tver, Russia}

\date{\today}
\def\be{\begin{equation}}
\def\ee{\end{equation}}
\def\bea{\begin{eqnarray}}
\def\eea{\end{eqnarray}}
\def\s{\sigma}
\def\al{\alpha}
\def\lm{\lambda}
\def\de{\delta}
\def\om{\omega}
\def\pr{\prime}
\def\f{\varphi}

\begin{abstract}
We consider various hadron models with a string carrying $n=3$
massive points (quarks): Y configuration, linear baryon model
$q$-$q$-$q$ and the closed string. For these models classical
rotational states (planar uniform rotations) are tested for
stability with respect to small disturbances. It is shown that
rotations of all mentioned models are unstable, but nature of this
instability is different. For the model Y the instability results
from existence of multiple real frequencies in the spectrum of
small disturbances, but for the linear model and the closed string
the similar spectra contain complex frequencies, corresponding to
exponentially growing modes of disturbances. This classical
rotational instability is important for describing excited
hadrons, in particular, for the linear  model and the closed
string it results in additional contribution in width of hadron
states.

\end{abstract}

\maketitle

\section{Introduction}\label{Intr}

\begin{figure}[bh]
\includegraphics[scale=0.9,trim=17 30 180 5]{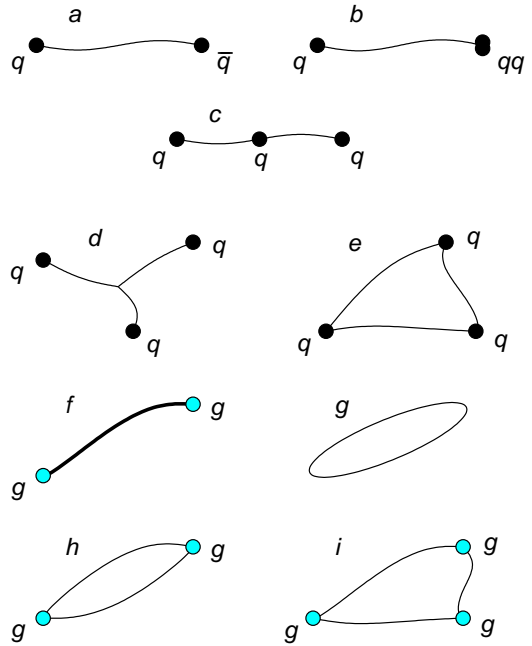}
\caption{String models of mesons, baryons and glueballs}
\label{mod}\end{figure}

In various string models of hadrons material points representing
quarks are connected by the Nambu-Goto strings (relativistic
strings) simulating strong interaction between quarks at large
distances
\cite{Nambu,Ch,AY,4B,Ko,lin,PY,PRTr,InSh,Solovm,stab,Y02}
(Fig.~\ref{mod}). This string has linearly growing energy (energy
density is equal to the string tension $\gamma$),  describes
contribution of the gluon field, QCD confinement mechanism and
quasilinear Regge trajectories for excited states of mesons and
baryons \cite{4B,Ko,lin,PY,PRTr,InSh,Solovm}.

Such a string with massive ends in Fig.~\ref{mod}{\it a} may be
regarded as the meson string model \cite{Ch}. String models of the
baryon were suggested in the following four topologically
different variants \cite{AY}: ({\it b}) the quark-diquark model
$q$-$qq$ \cite{Ko} $\big[$on the classic level it coincides with
the meson model ({\it a})$\big]$, ({\it c}) the linear
configuration $q$-$q$-$q$ \cite{lin}, ({\it d}) the
``three-string'' model or Y configuration \cite{AY,PY}, and (e)
the ``triangle'' model or $\Delta$ configuration \cite{PRTr}.
String models of the glueball \cite{Solovgl,EstrCBRS,ZayasS,glY08}
were considered in the variants in  Fig.~\ref{mod}{\it f\,--\,i}.
Here massive points describe valent gluons.

One should choose the most preferable model among the mentioned
four string baryon models in Fig.~\ref{mod}{\it b\,--\,e}. The
problem of this choice remains open
\cite{4B,Ko,lin,PY,PRTr,InSh,Solovm,stab,Y02}. Different models
have different advantages. In particular, rotational states
(planar uniform rotations) of all mentioned baryon models generate
linear or quasilinear Regge trajectories, but with different
slopes $\al'$ \cite{4B,Ko}. For the baryon models in
Fig.~\ref{mod}{\it b} and {\it c} like for the meson model in
Fig.~\ref{mod}{\it a} this slope and the string tension $\gamma$
are connected by the Nambu relation \cite{Nambu}
$\alpha'=1/(2\pi\gamma) $. The experimental value of this slope
$\al'\simeq0{.}9$ GeV$^{-2}$ is equal for both meson and baryon
Regge trajectories. So it is the argument in favour of these two
baryon models.

 For  rotational states of the linear baryon configuration
(Fig.~\ref{mod}{\it c}) the middle mass is at the rotational
center. In papers \cite{lin,stab} we have shown in numerical
experiments that the mentioned states are unstable with respect to
small disturbances and in Ref.~\cite{Unst09} we proved this result
analytically.

 The string baryon model Y (Fig.\,\ref{mod}{\it
d}) for its rotational states demonstrates Regge trajectories with
the slope \cite{PY} $\alpha'=1/(3\pi\gamma)$. To obtain
$\al'\simeq0{.}9$ GeV$^{-2}$  we are to assume that the effective
string tension $\gamma_Y$ in this model differs from $\gamma$ in
models in Figs.\,\ref{mod}{\it a\,--\,c} (the fundamental string
tension) and equals $\gamma_Y=\frac23\gamma$ \cite{4B,InSh}.
 Moreover, rotations of the Y string configuration are also unstable
with respect to small disturbances on the classic level
\cite{stab,Y02}. But specific features of this instability require
more profound investigation: in Sect.~\ref{Y} of this paper we
study difference in character of rotational instability for Y and
linear configurations.

The string baryon model ``triangle'' or $\Delta$  generates a set
of rotational states with different topology \cite{PRTr}. The so
called triangle states was applied for describing excited baryon
states on the Regge trajectories \cite{4B,InSh}, but in this case
(like for the model Y) we are to take another effective string
tension $\gamma_\Delta=\frac38\gamma$.

Different string models shown in Fig.~\ref{mod}{\it f\,--\,i} were
used for describing glueballs (bound states of gluons) and other
exotic hadrons \cite{Solovgl,EstrCBRS,ZayasS,glY08} predicted in
QCD. String models of glueballs include the open string with
enhanced tension $\gamma_{adj}=\frac94\gamma $
 (the adjoint string) and two constituent gluons at the ends
\cite{EstrCBRS,Bali0} in Fig.~\ref{mod}{\it f}; the closed string
without masses (Fig.~\ref{mod}{\it g})
\cite{Solovgl,ZayasS,MeyerT} and the closed string carrying
massive points (Fig.~\ref{mod}{\it h} and {\it i}) \cite{glY08}.

The problem of stability for rotations with respect to small
disturbances is very important for choosing the most adequate
string model for baryons or glueballs \cite{4B,stab,Y02,glY08}.
Note that instability of classical rotations for some string
configuration does not mean that the considered string model must
be totally prohibited. All excited hadron states (objects of
modelling) are resonances, they are unstable with respect to
strong decays. So they have rather large width $\Gamma$. On the
level of string models these decays are described as string
breaking with probability, proportional to the string length
$\ell$ \cite{Kodecay,GuptaR}. The corresponding width
$\Gamma=\Gamma_{br}\sim\ell$.

If classical rotations of a string configuration are unstable,
this instability gives the additional contribution $\Gamma_{inst}$
to width $\Gamma$. This effect is one of manifestations of
rotational instability. It can restrict applicability of some
string models, if the total width $\Gamma$ predicted by this model
(below we suppose $\Gamma =\Gamma_{br}+\Gamma_{inst}$) essentially
exceeds experimental data.

The stability problem for rotational states is solved for the
string with massive ends (Fig.~\ref{mod}{\it a, b}). Analytical
investigation of small disturbances demonstrated that rotational
states of this system are stable, and there is the spectrum of
quasirotational states in the linear vicinity of these stable
rotations \cite{stab,qrottmf}.

For string baryon models $q$-$q$-$q$, Y and $\Delta$ evolution of
small disturbances of rotational states was investigated in
numerical experiments \cite{stab,Y02}. These calculations
demonstrated instability of rotations for the linear model and for
the Y-configuration. However, we are to estimate analytically
increments of instability for all models and to investigate its
influence on properties of excited hadrons.

In this paper dynamics of the mentioned string models is described
in Sect.~\ref{Dyn}. In Sections \ref{Y} and \ref{Linear}  for the
models Y and $q$-$q$-$q$ correspondingly (Figs.~\ref{mod}{\it d}
and {\it c}) small disturbances of rotations are studied
analytically and increments of instability are calculated. In
Sect.~\ref{Closed} the similar result is presented for central
rotational states (with a massive point at the rotational center)
of the closed string (Fig.~\ref{mod}{\it e, h} or {\it i}). In
Sect.~\ref{Width} we study how  rotational instability enlarges
width of excited hadrons on Regge trajectories.

\section{Dynamics of a string with massive points}\label{Dyn}

Dynamics of an open or closed string carrying $n$ point-like
masses $m_1,\dots,m_n$ is determined by the action
\cite{PRTr,glY08}
 \be
A=-\gamma\int\limits_{D}\sqrt{-g}\;d\tau d\s -\sum\limits_{j=1}^n
m_j\int\sqrt{\dot x_j^2(\tau)}\;d\tau.
 \label{S}\ee

Here $\gamma$ is the string tension, $g$ is the determinant of the
induced metric
$g_{ab}=\eta_{\mu\nu}\partial_aX^\mu\partial_bX^\nu$ on the string
world surface $X^\mu(\tau,\s)$ embedded in Minkowski space $R^{1,3}$,
$\eta_{\mu\nu}=\mbox{diag}(1,-1.-1,-1)$, the speed of light $c=1$.

A world surface of the closed string mapping into $R^{1,3}$ from
the domain
 $$
 D=\big\{(\tau,\s):\,\tau\in R,\;\s_0(\tau)<\s<\s_n(\tau)\big\}
 $$
 is divided into $n$ world sheets by the world lines of massive points
$$x_j^\mu(\tau)=X^\mu(\tau,\s_j(\tau)),\quad j=0,1,\dots,n.$$
 Two of these functions
$x_0(\tau)$ and $x_n(\tau)$ describe the same trajectory of the
$n$-th massive point, and their equality forms the closure
condition
 \be
X^\mu(\tau,\s_0(\tau))=X^\mu(\tau^*,\s_n(\tau^*))
 \label{clos}\ee
 on the tube-like world surface \cite{PRTr,MilSh}. These equations may
contain two different parameters $\tau$ and $\tau^*$, connected
via the relation $\tau^*=\tau^*(\tau)$. This relation should be
included in the closure condition (\ref{clos}).

For the string baryon model $q$-$q$-$q$ (an open string with $n=3$
masses) the domain $D$ in Eq.~(\ref{S}) has the form
$\s_1(\tau)<\s<\s_3(\tau)$. This domain and the world surface are
divided into two sheets by the line $\s=\s_2(\tau)$. Naturally,
there is no closure condition in this model.

Equations of motion for both open and closed strings with massive
points result from the action (\ref{S}) and its variation. If we
use invariance of the action (\ref{S}) with respect to
nondegenerate reparametrizations $\tau=\tau(\tilde\tau,\tilde\s)$,
$\s=\s(\tilde\tau,\tilde\s)$ and choose the coordinates $\tau$,
$\s$ satisfying  the orthonormality conditions on the world
surface
 \be
(\partial_\tau X\pm\partial_\s X)^2=0,
 \label{ort}\ee
 the equations of motion are reduced to the
simplest form  \cite{4B,PRTr}. They include the string motion
equation
 \be \frac{\partial^2X^\mu}{\partial\tau^2}-
\frac{\partial^2X^\mu}{\partial\s^2}=0,
 \label{eq}\ee
 and equations for two types of massive points: for endpoints of
 the model $q$-$q$-$q$
 \bea
 m_1\frac d{d\tau}\frac{\dot
x_1^\mu(\tau)}{\sqrt{\dot x_1^2(\tau)}}-
\gamma\bigl[X^{'\!\mu}+\dot\s_1(\tau)\,\dot
X^\mu\bigr]\bigg|_{\s=\s_1}&=&0,\qquad \label{qq1}\\
m_3\frac d{d\tau}\frac{\dot x_3^\mu(\tau)}{\sqrt{\dot
x_3^2(\tau)}}+\gamma\bigl[X^{'\!\mu}+\dot\s_3(\tau)\,\dot
X^\mu\bigr]\bigg|_{\s=\s_3}&=&0;\qquad
 \label{qq3} \eea
 and for the middle point in the mentioned model or points on a
 closed string
 \bea m_j\frac d{d\tau}\frac{\dot
x_j^\mu(\tau)}{\sqrt{\dot x_j^2(\tau)}}+\gamma
\Big[X^{'\!\mu}+\dot\s_j(\tau)\dot
X^\mu\Big]\Big|_{\s=\s_j-0}&&\nonumber\\
{}-\gamma\Big[X^{'\!\mu}+\dot\s_j(\tau)\dot
X^\mu\Big]\Big|_{\s=\s_j+0}=0,&&
 \label{qqi}\eea
 \be  m_n\frac d{d\tau}\frac{\dot x_0^\mu(\tau)}{\sqrt{\dot
x_0^2(\tau)}}+\gamma
\big[X^{'\!\mu}(\tau^*,\sigma_n)-X^{'\!\mu}(\tau,0)\big]=0.
\label{qq0} \ee
 Here $\dot X^\mu\equiv\partial_\tau X^\mu$,
$X^{'\!\mu}\equiv\partial_\s X^\mu$, the scalar product
$(\xi,\zeta)=\eta_{\mu\nu}\xi^\mu\zeta^\nu$.

In Eq.~(\ref{qq0}) for $n$-th massive point we fix
 \be \s_0(\tau)=0,\qquad \s_n(\tau)=2\pi
\label{s02pi}\ee
 without loss of generality with the help of substitutions
$\tau\pm\s=f_\pm(\tilde\tau\pm\tilde\s)$, keeping conditions
(\ref{ort}) (conformal flatness of the induced metric $g_{ab}$)
\cite{PRTr,MilSh}.

For the open string model $q$-$q$-$q$ we can fix the similar
conditions at the ends \cite{lin,4B} in Eqs.~(\ref{qq1}),
(\ref{qq3}):
 \be \s_1(\tau)=0,\qquad \s_3(\tau)=\pi.
\label{s0pi}\ee

\smallskip

For the string baryon model Y (Fig.\,\ref{mod}{\it d}) three world
sheets (swept up by three string segments) are parametrized with
three different functions $X_j^\mu(\tau_j,\s)$ \cite{stab,Y02}.
Here we use different notations $\tau_1$, $\tau_2$, $\tau_3$ for
``time-like" parameters and the same symbol $\s$ for ``space-like"
parameters. These three world sheets are joined along the world
line of the junction that may be set as $\s=0$ for all sheets
without loss of generality, so the action of this configuration
takes the form \cite{stab,Y02}
 \be A=-\sum_{j=1}^3\int
d\tau_j\bigg[\gamma\!\!\int\limits_0^{\s_j(\tau_j)}
\!\!\sqrt{-g_j}\,d\s+m_j\sqrt{\dot x_j^2(\tau_j)}\,\bigg].
\label{SY}\ee
 Here $g_j=\dot X_j^2X_j^{'\!2}-(\dot X_j,X_j^{'})^2$,
$\dot x_j^\mu(\tau_j)=\frac d{d\tau_j}X^\mu(\tau_j,\s_j)$, $\dot
X_j^\mu=\partial_{\tau_j}X_j^\mu$, other notations are the same.

At the junction of three world sheets $X_j^\mu(\tau_j,\s)$ the
parameters $\tau_j$ are connected as follows \cite{Y02}
$$\tau_2=\tau_2(\tau),\quad\tau_3=\tau_3(\tau),\quad\tau_1\equiv\tau.$$
So the condition in the junction takes the form
 \be
X_1^\mu\big(\tau,0\big)=X_2^\mu\big(\tau_2(\tau),0\big)=
X_3^\mu\big(\tau_3(\tau),0\big). \label{junc}\ee

Dynamical equations for the Y configuration result from the action
(\ref{SY}) and under the orthonormality conditions (\ref{ort})
$$
 (\partial_{\tau_j} X_j\pm\partial_\s X_j)^2=0, \qquad j=1,2,3
$$
and condition (\ref{s0pi}) $0\le\s\le\pi$ on three world sheets
take the form \cite{Y02}
\begin{eqnarray}
&&\frac{\partial^2X_j^\mu}{\partial\tau_j^2}-
\frac{\partial^2X_j^\mu}{\partial\s^2}=0, \label{eqy}\\
&&\sum_{j=1}^3 X_j^{'\!\mu}\big(\tau_j(\tau),0\big)\,
\dot\tau_j(\tau)=0,\label{qy}\\
&&m_j\frac {dU^\mu_j(\tau_j)}{d\tau_j}+\gamma
X_j^{'\!\mu}(\tau_j,\pi)=0. \label{qqy}\end{eqnarray}
 Here $\dot\tau_j =\frac d{d\tau}\tau_j(\tau)$,
 \be
 U^\mu_j(\tau_j)=\frac{\dot x^\mu_j(\tau_j)}{\sqrt{\dot
x^2_j(\tau_j)}}.
 \label{Uj}\ee

Equations (\ref{junc})\,--\,(\ref{qqy}) describe all motions of
the Y configuration like Eqs.~(\ref{clos})\,--\,(\ref{eq}),
(\ref{qqi}), (\ref{qq0}) for the closed string with masses and
Eqs.~(\ref{ort})\,--\,(\ref{qq3}), (\ref{qqi}) ($j=2$) for the
string baryon model $q$-$q$-$q$.

\section{Rotational stability for Y configuration}\label{Y}

Rotational states of the  Y configuration (Fig.\,\ref{mod}{\it d})
correspond to planar uniform rotation of three rectangular string
segments connected at the junction at angles of $120^\circ$
\cite{4B,Ko,PY}. These states may be parametrized as \cite{Y02}
 \be
 \underline X_j^\mu(\tau_j,\s)= \Omega^{-1}\big[\om\tau_j
e_0^\mu+\sin(\om\s)\cdot e^\mu(\tau_j+\Delta_j)\big].
 \label{roty}\ee
  Here $\tau_1=\tau_2=\tau_3$,  $e_0,\,e_1,\,e_2,\,e_3$ is the
orthonormal tetrade in Minkowski space $R^{1,3}$,
$\Delta_j=2\pi(j-1)/(3\om)$,
 \be
e^\mu(\tau)=e_1^\mu\cos\om\tau+e_2^\mu\sin\om\tau \label{e}\ee
 is the unit space-like rotating vector directed along the first
string segment. Below we consider the case \cite{Y02}
 \be
 m_1=m_2=m_3,\qquad v_1=v_2=v_3
 \label{m123}\ee

Expression (\ref{roty}) satisfies Eq.~(\ref{eqy}) and conditions
(\ref{ort}), (\ref{junc}), (\ref{qy}), (\ref{qqy}), if angular
velocity $\Omega$, the value $\om$, constant velocities $v_j$ of
the massive points are connected by the relations \cite{4B}
 \be
v_j=\sin(\pi\om)=\left[\Big(\frac{\Omega
m_j}{2\gamma}\Big)^2+1\right]^{1/2}- \frac{\Omega m_j}{2\gamma}.
\label{v1}\ee

In Refs.~\cite{stab,Y02} we demonstrated in numerical experiments,
that rotational states (\ref{roty}) of the Y configuration are
unstable with respect to small disturbances. Here we solve this
problem analytically.

Let us consider a slightly disturbed motion of this configuration
with a world surface $X_j^\mu(\tau_j,\s)$ close to the surface
$\underline X_j^\mu(\tau_j,\s)$ of the rotational state
(\ref{roty}) (below we underline values, describing rotational
states). This disturbed motion is described by the general
solution of Eq.~(\ref{eqy})
 \be
 X_j^\mu(\tau_j,\s)=\frac1{2}\bigl[\Psi^\mu_{j+}(\tau_j+\s)+
\Psi^\mu_{j-}(\tau_j-\s)\bigr],
 \label{soly}\ee
 for every world sheet. Functions
$\Psi^\mu_{j\pm}(\tau)$ have isotropic derivatives
 \be
 \dot\Psi_{j+}^2=\dot\Psi_{j-}^2=0.
 \label{isotr}\ee
 as a consequence of the orthonormality conditions (\ref{ort}).

If we substitute Eq.~(\ref{soly}) into conditions (\ref{junc}),
(\ref{qy}) and (\ref{qqy}), they may be reduced to the form
\cite{Y02}
\begin{eqnarray}
& \Psi^\mu_{1+}(\tau)+\Psi^\mu_{1-}(\tau)=
 \Psi^\mu_{j+}(\tau_j)+\Psi^\mu_{j-}(\tau_j), &\label{jps1}\\
& \sum\limits_{j=1}^3\big[\dot\Psi^\mu_{j+}(\tau_j)-
\dot\Psi^\mu_{j-}(\tau_j)\big]\dot\tau_j(\tau)=0,&
 \label{jps2}\\
&\displaystyle\dot\Psi_{j\pm}^\mu(\tau_j\pm\pi)=
 \frac{m_j}{\gamma}\Big[\sqrt{-\dot
U_j^2 (\tau_j)}\,U_j^\mu(\tau_j)\mp\dot
U_j^\mu(\tau_j)\Big].\quad&
 \label{PsUy}
\end{eqnarray}

If we know velocities $U_j^\mu(\tau_j)$ of massive ends, we can
calculate functions $X_j^\mu(\tau_j,\s)$ for world sheets via
equations (\ref{PsUy}) and (\ref{soly}). So we search velocities
 $U_j^\mu$ for disturbed motion as small corrections to
vectors $\underline U_j^\mu$ for rotational states (\ref{roty}):
 \be
U_j^\mu(\tau_j)= \underline U_j^\mu(\tau_j)+u_j^\mu(\tau_j).
 \label{U+u}\ee
 We suppose disturbances $u_j^\mu(\tau_j)$ to be small
 ($|u_j^\mu|\ll1$),
so we omit squares of $u_j^\mu$ when we substitute the expression
(\ref{U+u}) into equalities $U_j^2=\underline U_j^2=1$, resulting
in relations
 \be
\big(\underline U_j(\tau_j),u_j(\tau_j)\big)=0,
 \label{Uu}\ee
and into dynamical equations (\ref{jps1})\,--\,(\ref{PsUy}).

The vectors $\underline U_j^\mu$ for rotational states
(\ref{roty}) in Eq.~(\ref{U+u}) are
 \be
 \underline U_j^\mu(\tau_j)= (1-v_j^2)^{-1/2} 
 \big[e_0^\mu+v_j\acute e^\mu(\tau_j+\Delta_j)
\big],\label{Uroty}\ee
 where rotating vector
 $$\acute e^\mu(\tau)=-e_1^\mu\sin(\om\tau)+e_2^\mu\cos(\om\tau),$$
 is orthogonal to the vector (\ref{e}).

We suppose that for a disturbed motions the ``time'' parameters
 \be
\tau_j(\tau)=\tau+\de_j(\tau),\quad j=2,3,\quad|\de_j(\tau)|\ll1,
 \label{taude}\ee
 have small deviations $\de_2(\tau)$ and $\de_3(\tau)$ from $\tau\equiv\tau_1$.

When we substitute disturbances (\ref{U+u}) and (\ref{taude}) into
expressions (\ref{PsUy}) and equations (\ref{jps1}) and
(\ref{jps2}) at the junction's world line, we get identities for
rotational terms (\ref{Uroty}) and (in the linear approximation
with respect to $u_j^\mu$, $\de_j$) the following system of vector
equations for these disturbances:
 \be
\begin{array}{c}
 \makebox[6cm]{$\sum\limits_\pm\Big[Qu_1^\mu(\pm)\pm\dot
u_1^\mu(\pm)+\underline
U_1^\mu(\pm)\big(e(\pm),\dot u_1(\pm)\big)\Big]=$}\\
\makebox[6cm]{$=\sum\limits_\pm\Big\{Qu_j^\mu(\pm)\pm\dot
u_j^\mu(\pm)+\underline U_j^\mu(\pm)\big(e(\pm_j),\dot
u_j(\pm)\big)+$}\\ \displaystyle
+\frac\gamma{m_1}\Big[\dot\de_j(\tau)\dot{\underline\Psi}_{j\pm}^\mu(\tau)+
\de_j(\tau)\ddot{\underline\Psi}_{j\pm}^\mu(\tau)\Big]\Big\},\\
 \displaystyle
 \sum\limits_{j=1}^3\sum\limits_\pm\Big\{\mp
 \frac\gamma{m_1}\Big[\dot\de_j\dot{\underline\Psi}_{j\pm}^\mu(\tau)+
\de_j\ddot{\underline\Psi}_{j\pm}^\mu(\tau)\Big]+\\
 \makebox[6cm]{$+ \dot u_j^\mu(\pm)\pm Qu_j^\mu(\pm)\pm\underline
U_j(\pm)\big(e(\pm_j),\dot u_j(\pm)\big)\Big\}=0.$}
 \end{array}
 \label{sysu}\ee
 Here $(\pm)\equiv(\tau\pm\pi)$,
 $(\pm_j)\equiv(\tau\pm\pi+\Delta_j)$, functions
 \be
 \dot{\underline\Psi}_{j\pm}^\mu(\tau)=\frac{m_1Q}{\gamma
 c_1} \big[e_0^\mu+v_1\acute e^\mu(\mp_j)\pm c_1e^\mu(\mp_j)
 \big]
 \label{Psrot}\ee
 correspond to rotational state (\ref{roty}),
 \be
Q=\om v_1/c_1,\qquad c_1=\cos(\pi\om)=\sqrt{1-v_1^2}.
 \label{Q}\ee
 One should add Eqs.~(\ref{Uu}) to the system (\ref{sysu}).

The similar equations in Ref.~\cite{Y02} were deduced with the
following mistake in corrections, connected with the ``times''
(\ref{taude}):
 $$\dot{\underline\Psi}_{j\pm}^\mu(\tau_j)\simeq\dot{\underline\Psi}_{j\pm}^\mu(\tau)+
 \ddot{\underline\Psi}_{j\pm}^\mu(\tau)\cdot\de_j(\tau).$$
 In particular, for vectors (\ref{e}) this correction doesn't reduce
 to the substitution $\tau\to\tau+{}$const made in Ref.~\cite{Y02}
 in the expression  $e^\mu(\tau_j)\simeq
e^\mu(\tau)+\om\acute e^\mu(\tau)\cdot\de_j(\tau)$. This
correction has the following true form:
 \be
e^\mu(\tau_j\mp\pi+\Delta_j)\simeq
e^\mu(\tau\mp\pi+\Delta_j)+\om\acute e^\mu(\mp_j)\cdot\de_j(\tau).
 \label{eco}\ee

In Ref.~\cite{Y02} we search solutions of the system (\ref{sysu})
in the form
 \begin{eqnarray}
u_j^\mu(\tau)&=&\big[A_j^0e_0^\mu+A_j^ze_3^\mu+c_1A_je^\mu(\tau+\Delta_j)+{}\nonumber\\
&+&v_1^{-1}A_j^0 \acute
e^\mu(\tau+\Delta_j)\big]\exp(-i\xi\tau), \label{uA}\\
 \de_j(\tau)&=&\de_j\exp(-i\xi\tau),\quad j=2,3
 \label{dej}
 \end{eqnarray}
 with small complex amplitudes $A_j^0$, $A_j$, $A_j^z$, $\de_j$.
Projections of these equations onto basis vectors $e_0^\mu$,
$e^\mu(\tau)$, $\acute e^\mu(\tau)$, $e_3^\mu$ form the system of
algebraic equations with respect to these amplitudes. Here the
expression (\ref{uA}) satisfy conditions (\ref{Uu}) due to the
factor $v_1^{-1}A_j^0$ at $\acute e^\mu$.

Projections of the mentioned equations onto the vector $e_3^\mu$
are
 $$\begin{array}{c}
 (\xi\tilde c +Q\tilde s)(A_1^z+A_2^z+A_3^z)=0,\\
 (\xi\tilde s-Q\tilde c)(A_1^z-A_j^z)=0,\quad j=2,3.\rule{0mm}{1.4em}
 \end{array}$$
 They don't depend on corrections of the type (\ref{eco}). Here
 $$  \tilde c=\cos\pi\xi,\qquad\tilde s=\sin\pi\xi.
 $$
 In Ref.~\cite{Y02} we obtained solutions of these equations,
describing 2 types of small oscillations of rotating Y
configuration (in $e_3$-direction). Corresponding frequencies
$\xi$ of these oscillations are roots of the equations
 \be
 \xi/Q=\cot\pi\xi,\qquad
\xi/Q=-\tan\pi\xi.
 \label{zfreqy}\ee
 All roots of Eqs.~(\ref{zfreqy}) are simple roots and real numbers,
therefore amplitudes of such fluctuations do not grow with growing
time $t$.

Small disturbances in the rotational ($e_1,e_2$) plane are
described by the system of 9 linear equations with 8 unknown
values $A_j^0$, $A_j$, $\de_j$. These equations are projections
(scalar products) of  the system (\ref{sysu}) onto 3 vectors
$e_0$, $e(\tau)$, $\acute e(\tau)$. Eight independent equations
among them are reduced to the form:
 \be
\begin{array}{c}
Q_c(A_j^0-A_1^0)+i\xi\tilde c
(A_j-A_1)=i\xi D_j, \\
2(\xi\tilde sA_1-iQ_sA_1^0)=(iQ_s+\epsilon_j\om_c)A_j^0-
(\xi+i\epsilon_j\om)\tilde sA_j,\rule{0mm}{1.2em}\\
2(\om_cA_1^0-i\om\tilde sA_1) =(iQ_s\epsilon_j-\om_c)A_j^0+
(i\om-\epsilon_j\xi)A_j,\rule{0mm}{1.2em}\\
iQ_s(A_1^0+A_2^0+A_3^0)=\xi\tilde s(A_1+A_2+A_3),\rule{0mm}{1.2em}\\
2(Q_cA_1^0+i\xi\tilde cA_1)=\rule{0mm}{1.2em}\\
=\sum\limits_{j=2}^3 \Big[(Q_c+i\epsilon_j\om_s)A_j^0+
(i\xi-\epsilon_j\om)(\tilde cA_j-D_j)\Big],\\
2(i\om_sA_1^0-\om\tilde cA_1)=\rule{0mm}{1.2em}\\
=\sum\limits_{j=2}^3 \Big[(i\om_s-\epsilon_jQ_c)A_j^0-
(i\epsilon_j\xi+\om)(\tilde cA_j-D_j)\Big].
\end{array}
\label{peq}\ee
 Here $\epsilon_j=(-1)^j\sqrt3$, $D_j=\de_jQ/c_1$,
  $j=2,3$,
 $$\begin{array}{c}
Q_c=Q(1+v_1^{-2})\tilde c-\xi\tilde s,\quad
Q_s=Q(1+v_1^{-2})\tilde s+\xi\tilde c,\\
 \om_c=\om\tilde c-\xi c_1v_1^{-1}\tilde s,\quad
\om_s=\om\tilde s+\xi c_1v_1^{-1}\tilde c.
 \end{array}$$

Nontrivial solutions of this system exist if and only if its
determinant equals zero. This equality after simplification is
reduced to the following equation:
 \be
(\xi^2-\om^2)(\om Q_c-\xi\om_s)(\om Q_s+\xi\om_c)=0.
 \label{pfr1}\ee
 It is equivalent to equalities $\xi=\pm\om$ and two
 equations
 \be
\frac{\xi^2-q}{2Q\xi}=-\tan\pi\xi, \qquad
\frac{\xi^2-q}{2Q\xi}=\cot\pi\xi.
 \label{pfreqy}\ee
 Here $q=Q^2(1+v_1^{-2})=\om^2(1+v_1^2)/(1-v_1^2)$.

\begin{figure}[th]
\includegraphics[scale=0.7,trim=15 0 10 -3]{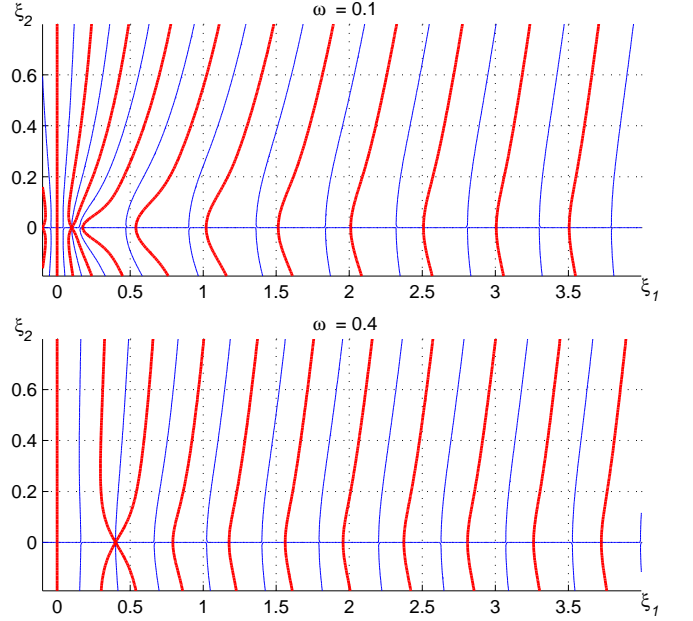}
\caption{Zero level lines for real (thick and) and imaginary
part(thin) of Eq.~(\ref{pfr1}) for specified values $\om$}
 \label{Y2}
 \end{figure}

Analysis of roots for equations (\ref{pfr1}), (\ref{pfreqy}) for
complex values $\xi=\xi_1+i\xi_2$ is presented in Fig.~\ref{Y2},
where the thick and thin lines are correspondingly zero level
lines of the real and imaginary part of the l.h.s. of
Eq.~(\ref{pfr1}) $f(\xi)=f(\xi_1+i\xi_2)$. Roots of this equation
are shown as cross points of a thick line with a thin line.

Fig.~\ref{Y2} shows that for values  $\om=0{.}1$ and $\om=0{.}4$,
corresponding to $v_1\simeq0.309$ and $v_1\simeq0.951$, all roots
of Eq.~(\ref{pfr1}) are real numbers and form a countable set. The
same behavior takes place for all values $\om\in(0,1/2)$, that is
for all rotations (\ref{roty}) with equal masses (\ref{m123}).

If a spectrum of small disturbances contains complex frequencies
$\xi=\xi_1+i\xi_2$, exponentially growing modes $|u_j|\sim
e^{\xi_2\tau}$ appear. Such a spectrum for rotations (\ref{roty})
of the model Y doesn't contain complex frequencies. So instability
of rotational states (\ref{roty}), observer in numerical
experiments \cite{stab,Y02}, has other nature.

This instability results from existence of double roots
$\xi=\pm\om$ in Eq.~(\ref{pfr1}).  If we take $\xi=\pm\om$ not
only the first factor in Eq.~(\ref{pfr1}), but also the second
factor $(\om Q_c-\xi\om_s)$ vanishes. This fact is seen in
Fig.~\ref{Y2} and may be proved, if we consider the first
Eq.~(\ref{pfreqy}) and Eqs.~(\ref{v1}), (\ref{Q}).

Double roots of Eq.~(\ref{pfr1}) may correspond to oscillatory
modes with linearly growing amplitude. To analyze this effect for
the frequency $\xi=\om$ we substitute small disturbances
\be\begin{array}{c}
 \!\!\!u_j^\mu(\tau)=\Big\{(A_j^0+\tilde A_j^0\tau)
 \big[e_0^\mu+v_1^{-1}\acute e^\mu(\tau+\Delta_j)\big]+\!\!\!\!\\
+(A_j+\tilde A_j\tau)\,c_1e^\mu(\tau+\Delta_j)\Big\}\exp(-i\om\tau), \label{}\\
 \de_j(\tau)=(D_j+\tilde D_j\tau)\,c_1Q^{-1}\exp(-i\om\tau) 
 \end{array} \label{uAdetau}
 \ee
 in the system (\ref{sysu}) for expressions (\ref{uA}) and (\ref{dej}).
This results after transformation in the following system
 \be\begin{array}{c}
 \tilde A_1^0=\tilde A_2^0=\tilde A_3^0=0, \\
 \tilde A_j=\frac12(i\epsilon_j-1)\tilde A_1,\quad
 \tilde D_j=\frac12c_1(i\epsilon_j-3)\tilde A_1,\rule{0mm}{1.4em}\\
 2A_j^0=iv_1c_1\om^{-1}\epsilon_j\tilde A_1-(i\epsilon_j+1)A_1^0,\rule{0mm}{1.4em}\\
 \makebox[6cm]{$\quad
 2A_j+(1-i\epsilon_j)A_1=\frac{i}{v_1c_1}\big[2A_j^0+(1-i\epsilon_j)A_1^0\big],
 \rule{0mm}{1.6em}
 $}\\
 2D_j=(3i+\epsilon_j)(\pi v_1\tilde A_1+v_1^{-1}A_1^0+ic_1A_1^0),\rule{0mm}{1.4em}
 \end{array} \label{systau}
 \ee
 with respect to complex amplitudes in (\ref{uAdetau}) (here
$j=2,3$). It is analog of Eqs.~(\ref{peq}).

The algebraic system (\ref{systau}) has a family of nontrivial
solutions specified with an arbitrary {\it nonzero} value of the
complex constant $\tilde A_1$. These solutions describe
disturbances (\ref{uAdetau}) of the rotational states with
linearly growing amplitude:
 \be
 |u_j^\mu|\simeq|\tilde A_1|c_1\tau\qquad
 |\de_j|\simeq \sqrt3c_1^2Q_1^{-1}|\tilde A_1|\tau. \label{linmod}
 \ee

This modes let us to conclude, that rotational states (\ref{roty})
of the string model Y with $m_1=m_2=m_3$ are unstable, because an
arbitrary small disturbance contains linearly growing modes of the
type (\ref{linmod}) in its spectrum.

In papers \cite{Y02,stab} we investigated numerically disturbed
rotational states of the string configuration Y and observed
instability of the states (\ref{roty}). Small disturbances grew
and resulted in transformation of the Y-shaped three-string into
the linear $q$-$q$-$q$ configuration after merging a massive point
with the junction.

\begin{figure}[th]
\includegraphics[scale=0.7,trim=15 2 10 3]{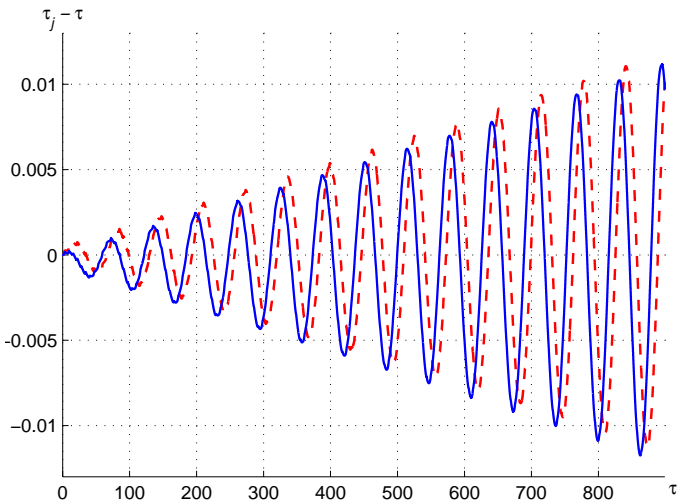}
\caption{Dependence of $\tau_2(\tau)-\tau$ (solid line) and
$\tau_3(\tau)-\tau$ (dashed line) on $\tau$ for disturbed
rotational state}
 \label{Ygr}\end{figure}

Numerical experiments demonstrate that evolution of small
disturbances for velocities $U_j^\mu$ or values $\tau_j(\tau)$
corresponds to expression (\ref{linmod}), amplitudes of
disturbances linearly grow and frequency of oscillations (with
respect to $\tau$) is equal $\om$. Omitting details of numerical
modelling for motions close to rotations (\ref{roty}) (described
in Refs.~\cite{Y02,stab}), we demonstrate in Fig.~\ref{Ygr}
dependence of deviations $\tau_2(\tau)-\tau$ (solid line) and
$\tau_3(\tau)-\tau$ (dashed line) on the time parameter $\tau$ for
disturbed rotational states (\ref{roty}).
 Here we test the state with masses (\ref{m123}) for $\om=0.1$ and
 with the initial disturbance of the component $\dot\Psi_{1\pm}^1(\tau)$ less than
0.01 of its value (\ref{Psrot}).

Instability of rotational states (\ref{roty}) with linearly
growing amplitudes takes place also in the limit $m_j\to0$, or, it
is equivalent, $v_j\to1$.

\section{Rotational states and their stability for linear model}\label{Linear}

 Rotational states of the linear string model $q$-$q$-$q$ are planar
uniform rotations of the rectilinear string segment with the middle quark at the
rotational center. These rotations may be described by the following exact solution of
equations (\ref{ort})\,--\,(\ref{qqi}) \cite{stab}:
 \be
\underline X^\mu(\tau,\s)=\Omega^{-1} \big[\om\tau
e_0^\mu+\cos(\om\s+\phi_1)\cdot e^\mu(\tau)\big],
 \label{lrot}\ee
 Here $\s\in[0,\pi]$, $\Omega$
is the angular velocity, $e^\mu(\tau)$ is the unit space-like
rotating vector (\ref{e}) directed along the string. Values $\om$
(dimensionless frequency) and $\phi_1$ are connected with the
constant speeds $v_j$ of the ends
 \be
  v_1=\cos\phi_1,\;\;\;  v_3=-\cos(\pi\om +\phi_1),\;\;\;
\frac{m_j\Omega}\gamma=\frac{1-v_j^2}{v_j},
 \label{v13}\ee
 where $j=1,3$. The
central massive point of the $q$-$q$-$q$ system is at rest (in the
corresponding frame of reference) at the rotational center. Its
inner coordinate is
 \be
\s_2(\tau)=\underline\s_2=\frac{\pi-2\phi_1}{2\om}={\mbox{const}}.
\label{sig2}\ee

Rotational states (\ref{lrot}) of the model $q$-$q$-$q$ was tested
for stability in Ref.~\cite{stab} in numerical experiments and in
Ref.~\cite{Unst09} analytically. These experiments and
calculations demonstrated instability (growth of small
disturbances). Here we use analytical approach  applied in
Refs.~\cite{qrottmf,Unst09} for testing stability of the states
(\ref{lrot}).

Let us consider a slightly disturbed motion of the system
$q$-$q$-$q$ in the linear vicinity of the rotational state
(\ref{lrot}). This disturbed motion is described by the general
solution (\ref{soly}) of Eq.~(\ref{eq})
 \be
X^\mu (\tau,\s)=\frac{1}{2}\big[\Psi^\mu_{j+}(\tau +\s)+\Psi
^\mu_{j-}(\tau-\s)\big].
 \label{gensol}\ee
 Here $j=1$ for  $\s\in[0,\s_2]$ and $j=2$ for  $\s\in[\s_2,\pi]$, functions
$\Psi^\mu_{j\pm}(\tau)$ have isotropic derivatives (\ref{isotr})
 due to the orthonormality conditions (\ref{ort}).
The  functions $\Psi^\mu_{j\pm}$ are smooth, the world surface
(\ref{gensol}) (smooth if $\s\ne\s_2$) is continuous at the line
$\s=\s_2(\tau)$. This condition in terms Eq.~(\ref{gensol}) takes
the form
 \be
\Psi^\mu_{1+}(+_2)+\Psi^\mu_{1-}(-_2)=
\Psi^\mu_{2+}(+_2)+\Psi^\mu_{2-}(-_2),
 \label{cont2}\ee
 where $(\pm_2)\equiv\big(\tau\pm\s_2(\tau)\big)$.

We use underlined symbols for describing the particular exact
solution (\ref{lrot}) for the rotational states. For example, we
denote
 \be
\underline\Psi^{\mu}_{1\pm}(\tau)=\underline\Psi^\mu_{2\pm}(\tau)=
\Omega^{-1}\big[e_0^\mu\om\tau+e^\mu(\tau\pm\phi_1/\om)\big]
 \label{Pslin}\ee
 the functions in Eq.~(\ref{gensol}) corresponding to the
 states (\ref{lrot}):
 $\underline X^\mu=\frac{1}{2}\big[\underline\Psi^\mu_{j+}(\tau +\s)+
 \underline\Psi^\mu_{j-}(\tau-\s)\big]$.

To describe any small disturbances of the rotational motion, that
is motions close to states (\ref{lrot}) we consider vector
functions $\Psi^\mu_{j\pm}$ close to $\underline\Psi^\mu_{j\pm}$
(\ref{Pslin}) in the form
 \be
\Psi^\mu_{j\pm}(\tau)=\underline\Psi^{\mu}_{j\pm}(\tau)
+\psi_{j\pm}^\mu(\tau).\label{Psips}\ee

Disturbances $\psi_{j\pm}^\mu(\tau)$ are supposed to be small, so
we omit squares of $\psi_{j\pm}$ when we substitute the expression
(\ref{Psips}) into dynamical equations
(\ref{qq1})\,--\,(\ref{qqi}) and (\ref{cont2}). In other words, we
work in the first linear vicinity of the states (\ref{lrot}). Both
functions $\dot\Psi^\mu_{j\pm}$ and
$\dot{\underline\Psi}^\mu_{j\pm}$ in expression (\ref{Psips}) must
satisfy the condition (\ref{isotr})
 resulting from Eq.~(\ref{ort}), hence in the first order approximation on $\dot\psi_{j\pm}$
the following scalar product equals zero:
 \be
\big(\underline{\dot\Psi}_{j\pm},\dot\psi_{j\pm}\big)=0.
 \label{scPs}\ee

For disturbed (quasirotational) motions of the model $q$-$q$-$q$
the inner coordinate $\s_2(\tau)$ of the middle massive point
differs from the constant value $\underline\s_2$ (\ref{sig2}) and
should include the following small correction $\de_2$:
 \be
\s_2(\tau)=\underline\s_2+\de_2(\tau).
 \label{s2d}\ee

If we substitute expressions (\ref{Psips}), (\ref{s2d}) with
(\ref{gensol}) into the continuity condition (\ref{cont2}) and
three equations (\ref{qq1}), (\ref{qq3}), (\ref{qqi}) (with $j=2$)
for massive points, we obtain equalities for summands with
$\underline\Psi^\mu_{j\pm}$ and four equations for small
disturbances $\psi_{j\pm}^\mu(\tau)$  in the first linear
approximation:
 \be\begin{array}{c}
\psi_{1+}^\mu(+_2)+\psi_{1-}^\mu(-_2)=
\psi_{2+}^\mu(+_2)+\psi_{2-}^\mu(-_2),\\
\dot\psi_{1+}^\mu+\dot\psi_{1-}^\mu-\underline
U_1^\mu\big(\underline
U_1,\dot\psi_{1+}^\mu+\dot\psi_{1-}^\mu\big)=Q_1\big(\psi_{1+}^\mu-\psi_{1-}^\mu\big),
\rule{0mm}{1.2em}\\
\dot\psi_{2+}^\mu(+)+\dot\psi_{2-}^\mu(-)-\underline
U_3^\mu\big(\underline
U_3,\dot\psi_{2+}^\mu(+)+\dot\psi_{2-}^\mu(+)\big)\rule{0mm}{1.2em}\\
 =Q_3\big[\psi_{2-}^\mu(-)-\psi_{2+}^\mu(+)\big],
\rule{0mm}{1.2em}\\
\dot\psi_{1+}^\mu(+_2)+\dot\psi_{1-}^\mu(-_2)-2a_0
\big[\dot\de_2e^\mu(\tau)+\om\de_2\acute e^\mu(\tau)\big]\rule{0mm}{1.2em}\\
 \displaystyle
=\frac{e_0^\mu}{2a_0}\Big[\big(\dot{\underline\Psi}_{1+}(+_2),\dot\psi_{1-}(-_2)\big)+
\big(\dot{\underline\Psi}_{1-}(-_2),\dot\psi_{1+}(+_2)\big)\Big]\rule{0mm}{1.7em}\\
 +2Q_2\big[\psi_{2+}^\mu(+_2)-\psi_{1+}^\mu(+_2)\big].\rule{0mm}{1.2em}
\end{array}
 \label{syslin}\ee
 Here
 \be
 Q_j=\frac\gamma{m_j}\sqrt{\underline{\dot x}_j^2(\tau)}=
 \frac{\gamma a_0}{m_j}\sqrt{1-v_j^2},\quad
a_0=\frac\om\Omega,
 \label{Qj}\ee
 vector-functions similar to (\ref{Uroty})
$$
\underline
U_j^\mu(\tau)=(1-v_j^2)^{-1/2}\big[e_0^\mu-\epsilon_jv_j\acute
e^\mu(\tau)\big],\quad\epsilon_1=-1,\quad\epsilon_3=1,
$$
 are unit velocity vectors of the moving massive points.

If we consider projections (scalar products) of 4 equations
(\ref{syslin}) onto 4 basic vectors $e_0$, $e(\tau)$, $\acute
e(\tau)$, $e_3$ and add Eqs.~(\ref{scPs}) we obtain the system of
20 differential equations with deviating arguments with respect to
17 unknown functions: $\de_2(\tau)$ and 16 projections
 \be \begin{array}{c}
  \psi_{j\pm}^0\equiv( e_0,\psi_{j\pm}),\quad
\psi_{j\pm}^3\equiv( e_3,\psi_{j\pm}),\\
 \psi_{j\pm}\equiv ( e,\psi_{j\pm}), \quad
\acute\psi_{j\pm}\equiv(\acute e,\psi_{j\pm});
 \rule{0mm}{1.2em}\end{array}
 \label{psiscal}\ee

Four projections of Eqs.~(\ref{syslin}) onto direction $e_3$
(orthogonal to the rotational plane $e_1$,\,$e_2$) form the closed
subsystem with respect to 4 functions (\ref{psiscal})
$\psi_{j\pm}^3$:
 \be
\!\!\begin {array}{c} \psi_{1+}^3(+_2)+\psi_{1-}^3(-_2)=
\psi_{2+}^3(+_2)+\psi_{2-}^3(-_2),\\
\dot\psi_{1+}^3(\tau)+\dot\psi_{1-}^3(\tau)=
Q_1\big[\psi_{1+}^3(\tau)-\psi_{1-}^3(\tau)\big],\rule{0mm}{1.2em}\\
\dot\psi_{2+}^3(+)+\dot\psi_{2-}^3(-)=
Q_3\big[\psi_{2-}^3(-)-\psi_{2+}^3(+)\big],\rule{0mm}{1.2em}\\
 \dot\psi_{1+}^3(+_2)+\dot\psi_{1-}^3(-_2)=2Q_2\big[
\psi_{2+}^3(+_2)-\psi_{1+}^3(+_2)\big].
\rule{0mm}{1.2em}\end{array}\!\!\!\!\!
 \label{sysps3}\ee

We search solutions of this homogeneous system  in the form of
harmonics similar to (\ref{uA})
 \be
\psi^3_{j\pm}=B_{j\pm}^3 e^{-i\xi\tau}.
 \label{psexp3} \ee

This substitution results in the linear homogeneous system of 4
algebraic equations with respect to 4 amplitudes $B_{j\pm}^3$. The
system has nontrivial solutions if and only if its determinant
equals zero:
 $$
 \left|\begin {array}{cccc}
 i\xi+Q_1 & i\xi-Q_1 & 0 & 0\\
 0 & 0 & (i\xi-Q_3)\,e^{-2i\pi\xi} & i\xi+Q_3 \\
 i\xi-2Q_2 & i\xi e^{2i\underline\s_2\xi} &
 -2Q_2 & 0\\
 e^{-i\underline\s_2\xi} &  e^{i\underline\s_2\xi} &-e^{-i\underline\s_2\xi}
 &- e^{i\underline\s_2\xi}
 \end{array}\right|=0.
 $$
  This equation is reduced to
 the form
  \bea
&Q_2\big[(Q_1Q_3-\xi^2)\sin\pi\xi+(Q_1+Q_3)\,\xi\cos\pi\xi\big]&\nonumber\\
&{}+\xi(Q_1\tilde c_1-\xi \tilde s_1)(Q_3\tilde c_3-\xi \tilde
s_3)=0&
 \label{lfreq3}\eea
 where
 $$\begin{array}{c}\tilde c_1=\cos\underline\s_2\xi,\quad\tilde
 s_1=\sin\underline\s_2\xi,\\
 \tilde c_3=\cos(\pi-\underline\s_2)\xi,\quad \tilde s_3=\sin(\pi-\underline\s_2)\xi.
 \end{array}$$

  The
spectrum of transversal (with respect to the $e_1,\,e_2$ plane)
small fluctuations of the string for the considered rotational
state contains frequencies $\xi$ which are roots of
Eq.~(\ref{lfreq3}). We search complex roots $\xi=\xi_1+i\xi_2$ of
this equation.

\begin{figure}[bh]
\includegraphics[scale=0.8,trim=13mm 2mm 15mm 2mm]{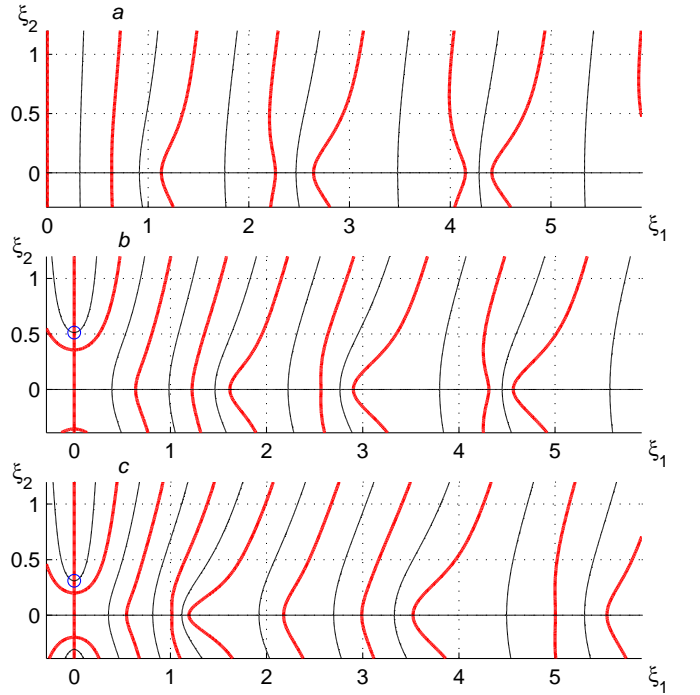}
\caption{Zero level lines for real part (thick) and imaginary part
(thin) {\it (a)} for  Eq.~(\ref{lfreq3}) with $Q_1=Q_2=Q_3=1$;
{\it (b)} for Eq.~(\ref{lfreq}) with $Q_1=Q_2=Q_3=1$; {\it (c)}
 for Eq.~(\ref{lfreq}), $Q_1=1$, $Q_2=0{.}2$, $Q_3=0{.}4$ }
 \label{frl}\end{figure}

In Fig.~\ref{frl}{\it a} the thick and thin lines are zero level
lines correspondingly present real and imaginary part of the
l.h.s. $f(\xi)=f(\xi_1+i\xi_2)$ of Eq.~(\ref{lfreq3}) for given
values $Q_j$. Roots of this equation are shown as cross points of
a thick line with a thin line. If the values (\ref{Qj}) $Q_j$ are
given, one can determine values $\om$, $\underline\s_2$, $v_j$,
$m_j/\gamma$ from Eqs.~(\ref{v13}), (\ref{s2d}), (\ref{Qj}). In
particular, values $\om$, $Q_1$, $Q_3$ are connected by the
relation
 \be
\om(Q_1+Q_3)=(\om^2-Q_1Q_3)\tan\pi\om,
 \label{omQ}\ee
 resulting from the mentioned equations.

 Analysis of roots of Eq.~(\ref{lfreq3}) for various values $Q_j$,
$m_j$ and $v_j$ shows, that for all values of mentioned parameters
all these frequencies are real numbers (cross points lie on the
real axis), therefore amplitudes of such fluctuations do not grow
with growing time $t$.

Note that any complex frequency $\xi=\xi_1+i\xi_2$ with positive
imaginary part $\xi_2$ result in exponential growth of the
corresponding amplitude of disturbances
 $$
 \psi_{k}^3=B_{k}^3\exp(-i\xi_1\tau)\cdot\exp(\xi_2\tau)
 $$
 In this case the considered state will be unstable \cite{stab,Unst09}.

To study small disturbances in the $e_1,\,e_2$ plane we consider
projections (scalar products) of equations (\ref{syslin}) onto 3
vectors $e_0$, $e(\tau)$, $\acute e(\tau)$. They form the system
of 12 differential equations with deviating arguments with respect
to 9 unknown functions $\psi_{j\pm}$, $\acute\psi_{j\pm}$,
$\de_2$, if functions $\psi_{j\pm}^0$ are excluded via the
orthonormality condition, Eqs.~(\ref{scPs}):
 $$
\dot\psi_{j\pm}^0=\pm c_1 (e,\dot\psi_{j\pm})-v_1(\acute
e,\dot\psi_{j\pm}),\qquad c_1=\cos\underline\s_2\om.
 $$

Only 9 from these 12 equations are independent ones. When we
search solutions of this system in the form of harmonics
(\ref{psexp3})
 \be
\psi_{j\pm}=B_{j\pm} e^{-i\xi\tau},\quad \acute\psi_{j\pm}=\acute
B_{j\pm} e^{-i\xi\tau}, \quad 2a_0\de_2=\Delta_2 e^{-i\xi\tau},
 \label{fexp} \ee
 we obtain the homogeneous system of 9 algebraic equations with
respect to 9 amplitudes $B_{j\pm}$, $\acute B_{j\pm}$, $\Delta_2$
(it is convenient to use the linear combinations of them
$A_{j\pm}=-i\xi B_{j\pm}-\om\acute B_{j\pm}$, $\acute
A_{j\pm}=-i\xi\acute B_{j\pm}+\om B_{j\pm}$):
 $$
\begin{array}{c}
 K_1^+A_{1+}+K_1^-A_{1-}
 =\acute K_1^+\acute A_{1+}+\acute K_1^-\acute A_{1-},\\
 (1-i\xi Q_1^\xi)\,A_{1+}+(1+i\xi Q_1^\xi)\,A_{1-}=\om Q_1^\xi
 (\acute A_{1-}-\acute A_{1+}), \rule{0mm}{1.2em}\\
 (v_1\Gamma_1-\om Q_1^\xi)(A_{1-}-A_{1+})\rule{0mm}{1.2em}\\
 =(1-i\xi Q_1^\xi)\,\acute A_{1+}+(1+i\xi Q_1^\xi)\,\acute A_{1-},
 \rule{0mm}{1.2em}\\
 c_1(A_{1+}-A_{2+})-v_1(\acute A_{1+}-\acute A_{2+})=0,\rule{0mm}{1.2em}\\
 c_1(A_{1-}-A_{2-})+v_1(\acute A_{1-}-\acute A_{2-})=0,\rule{0mm}{1.2em}\\
 K_1^+E_3^+A_{2+}+K_1^-E_3^-A_{2-}=\acute K_1^+E_3^+\acute A_{2+}+
 \acute K_1^-E_3^-\acute A_{2-},
 \rule{0mm}{1.2em}\\
 K_2^+E_3^+A_{2+}+K_2^-E_3^-A_{2-}=\acute K_2^+E_3^+\acute A_{2+}+
 \acute K_2^-E_3^-\acute A_{2-},\rule{0mm}{1.2em}\\
 (c_1-2\om v_1Q_2^\xi)\,E_2^+A_{1+}+c_1E_2^-A_{1-}+v_1E_2^-\acute A_{1-}
 -i\xi\Delta_2\rule{0mm}{1.2em}\\
 =E_2^+\big[(v_1+2\om c_1Q_2^\xi)\,\acute A_{1+}-2\om Q_2^\xi(v_1 A_{2+}+
 c_1\acute A_{2+})\big], \rule{0mm}{1.2em}\\
 \acute K_1^+E_2^+A_{1+}+\acute K_1^-E_2^-A_{1-}+K_1^+E_2^+\acute A_{1+}+
 K_1^-E_2^-\acute A_{1-} \rule{0mm}{1.2em}\\
 =-(\xi^2+\om^2)\,\Delta_2. \rule{0mm}{1.2em}
 \end{array}
 $$
 Here $\;Q_j^\xi=Q_j/(\xi^2-\om^2)$, $E_j^\pm=\exp(\mp
 i\xi\underline\s_j)$,
 $$ \begin{array}{c}
 K_1^\pm=c_1\om \mp iv_1\xi,\qquad\acute K_1^\pm=\pm v_1\om+ic_1\xi,\\
 K_2^\pm=\om Q_3^\xi\sin\pi\om-(1\pm i\xi Q_3^\xi)\cos\pi\om,\\
 \acute K_2^\pm=\mp\om Q_3^\xi\cos\pi\om-(\pm1+ i\xi
 Q_3^\xi)\sin\pi\om.
 \end{array}
 $$

Nontrivial solutions of this system exist if the condition similar
to Eq.~(\ref{lfreq3}) takes place. It may be reduced to the
following equation
 \be
 \frac\xi{Q_2}\cdot\frac{\xi^2-\om^2}{\xi^2+\om^2}=
 \sum_{j=1,3}
\frac{(q_j-\xi^2)\,\tilde c_j-2Q_j\xi\tilde s_j}
 {(q_j-\xi^2)\,\tilde s_j+2Q_j\xi\tilde c_j}.
 \label{lfreq}\ee
 Here $q_j=Q_j^2(1+v_j^{-2})$.

Fig.~\ref{frl}{\it b, c} demonstrates roots $\xi=\xi_n$ of
Eq.~(\ref{lfreq}), corresponding to frequencies of small
oscillations of the rotating system $q$-$q$-$q$ in the rotational
plane. Unlike Eq.~(\ref{lfreq3}), describing oscillations in $z$-
or $e_3$-direction, equation (\ref{lfreq}) always has two
imaginary roots $\xi=\pm i\xi_2^*$. The positive imaginary roots
$\xi=i\xi_2^*$, $\xi_2^*>0$ are marked with a circle in
Fig.~\ref{frl}{\it b, c}.

 Other roots of Eq.~(\ref{lfreq}) are
real ones. In Figs.~\ref{frl}{\it b} and \ref{frl}{\it a} values
$Q_j$, $m_j$ are the same, the mass relation here is
$m_1:m_2:m_3\simeq1:1.85:1$; for the case in Fig.~\ref{frl}{\it c}
it is $m_1:m_2:m_3\simeq1:10.5:4.2$.

The positive imaginary root $\xi=i\xi_2^*$ of  Eq.~(\ref{lfreq})
may be found after substituting $\xi=i\xi_2$:
 $$
 \frac{\xi_2}{Q_2}\cdot\frac{\xi_2^2+\om^2}{\om^2-\xi_2^2}=
\sum_{j=1,3} \frac{q_j+\xi_2^2+2Q_j\xi_2\tanh\check\s_j\xi_2}
 {(q_j+\xi_2^2)\,\tanh\check\s_j\xi_2+2Q_j\xi_2}.
 $$
Here $\check\s_1=\underline\s_2$, $\check\s_3=\pi-\underline\s_2$.
Evidently, the required value $\xi^*$ exists in the interval
$(0,\om)$. An arbitrary disturbed motion of the $q$-$q$-$q$
configuration contains exponentially growing modes in its
spectrum, in particular,
 \be
 \psi_{j\pm}=B_{j\pm}\exp(\xi_2^*\tau).
 \label{expps}\ee
 So the rotational motion
(\ref{lrot}) is unstable with respect to small disturbances.
Evolution of this instability was numerically analysed in
Ref.~\cite{stab}.

\section{Rotational states for closed string}\label{Closed}

 For the case of closed string rotational states (planar uniform rotations of
the string with massive points) were described and classified in
Refs.~\cite{PRTr,glY08}. These states are divided into 3 classes
\cite{glY08}:
 ``hypocycloidal states'' (in this case segments of rotating string,
connecting massive points, are segments of a hypocycloid),
``linear states'' and ``central states'', describing rotating
folded closed string with rectilinear string segments. For linear
states all masses $m_j$ move at nonzero velocities $v_j$, but in
the case of central states a massive point (or some of them) is
placed at the rotational center.

In Ref.~\cite{Unst09} we solved the stability problem for the
central rotational states with $n=3$ massive points where the mass
$m_3$ is at the center and other masses $m_1$ and $m_2$ rotates at
the ends of rectilinear segments. These states look like the
states (\ref{lrot}) of the linear model $q$-$q$-$q$, but have the
additional string segment (the string is closed) and another
numeration of massive points.

The mentioned central states  with $n=3$ have the form
 \be
\underline X^\mu=e_0^\mu a_0\tau+ u(\s)\cdot e^\mu(\tau),
 \label{Xc}\ee
 where  $u(\sigma)=
A_j\cos\omega\sigma+ B_j\sin\omega\sigma$,
$\sigma\in[\sigma_{j-1},\sigma_j]$ is continuous function, but its
derivatives  has discontinuities at
$\s=\s_j\equiv\underline\s_j={}$const (positions of masses
 $m_j$).
 They are described by following parameters, determined by
Eqs.~(\ref{clos}), (\ref{ort}), (\ref{qqi}), (\ref{qq0}):
 $A_1=0$,
$\underline\s_2-\underline\s_1=\pi$,
 $A_2=2\breve S_1\breve C_1B_1$, $B_2=(\breve S_1^2-\breve C_1^2)B_1$, $A_3=-\breve
SB_1$, $B_3=\breve CB_1$; $v_1=\breve S_1$,
$v_2=\sin(2\pi-\underline\s_2)\om$. Here
 $$
\breve C_j=\cos\om\underline\s_j,\quad\breve
 S_j=\sin\om\underline\s_j,\quad
\breve C\equiv\breve C_3,\quad\breve S\equiv\breve S_3,$$
  the closure condition (\ref{clos}) takes the form
 $\tau^*=\tau$ the values (\ref{Qj})
 $\gamma m_j^{-1}\sqrt{\dot X^2(\tau,\underline\s_j)}=Q_j$
  are constants.

For these rotational states  the string rotates at the angular
velocity $\Omega=\om/a_0$, the value $a_0$ connected with speeds
$v_j$ of massive points by the following equations, resulting from
Eqs.~(\ref{Qj}):
 \be a_0
=\frac{m_1Q_1}{\gamma \sqrt{1-v_1^2}}=\dots=\frac{m_n Q_n}{\gamma
\sqrt{1-v_n^2}}.
 \label{a0v}\ee

The central rotational states (\ref{Xc}) were tested for stability
in Ref.~\cite{Unst09}. In this paper the approach suggested in
Sect.~\ref{Linear} for states (\ref{lrot}) was used. Here we omit
details of this investigation and present its results: the central
states (\ref{Xc}) appeared to be unstable (small disturbances
grow) if the central mass is less than the critical value
 \be
 0<m_3<m_{3cr},
 \label{m3cr}\ee
 $$
m_{3cr}=2\pi\gamma a_0+\frac{m_1}{\sqrt{1-v_1^2}}
+\frac{m_2}{\sqrt{1-v_2^2}}\equiv E-m_3.
 $$
 Here is energy of the state (\ref{Xc}).
 We may conclude, the central rotational state is unstable if the
central mass $m_3$ is nonzero and less than energy of the string
with other massive points. Note that in the case of the linear
string model in Sect.~\ref{Linear} there were no such a threshold
effect.

If $m_3$ satisfies the condition (\ref{m3cr}), the spectrum of
small disturbances of the central rotational states (\ref{Xc}) has
complex frequencies. This results in exponential growth of small
disturbances in accordance with the expression (\ref{expps}).

In the case $m_3=0$ there is no massive point at the center, and
the corresponding linear rotational state with $n=2$ is stable.
Stability also takes place for the case $m_3\to\infty$.

\section{Instability of rotational states and hadron's width}\label{Width}

Rotational states (\ref{roty}) of the string model Y and
(\ref{lrot}) of the linear string model were applied for
describing orbitally excited baryons \cite{4B,InSh}. The similar
states (\ref{Xc}) of the closed string describe the Pomeron
trajectory \cite{glY08}, corresponding to possible glueball
states.

For rotational states (\ref{roty}), (\ref{lrot}) and (\ref{Xc})
energy $E$ or mass $M$ and angular momentum $J$ are determined by
the following expressions \cite{4B,InSh,glY08}:
 \bea
M=E&=&q\pi\gamma  a_0+\sum_{j=1}^n
\frac{m_j}{\sqrt{1-v_j^2}}+\Delta
E_{SL}, \label{E}\\
 J=L+S&=&
\frac{a_0}{2\omega}\bigg(q\pi\gamma a_0+\sum_{j=1}^n
\frac{m_jv_j^2}{\sqrt{1-v_j^2}} \bigg)+\sum_{j=1}^ns_j.\qquad
 \label{J}\ \eea
 Here $q=1$ for the linear model, $q=2$ for the closed string,
 $q=3$ for the Y model,
$s_j$ are spin projections of massive points (quarks or valent
gluons), $\Delta E_{SL}$ is the spin-orbit contribution to the
energy in the form \cite{4B}
 $$
\Delta E_{SL}=\sum_{j=1}^n\big[1-(1-v_j^2)^{1/2}\big] (
\textbf{$\Omega$}\cdot \textbf{s}_j).
 $$

If the string tension $\gamma$, values $m_j$ and $s_j$  are fixed,
we obtain the one-parameter set of rotational states (\ref{roty}),
(\ref{lrot}) or (\ref{Xc}). Values $J$ and $E^2$ for these states
form the quasilinear Regge trajectory with asymptotic behavior
 $
 J\simeq\al_0+\al'E^2,
 $
 for large $E$ and $J$ \cite{4B,glY08} with the slope
$\al'=1/(2\pi\gamma)$ for the linear quark-diquark models,
$\alpha'=1/(3\pi\gamma)$ for states (\ref{roty}) of the Y model
and $\alpha'=1/(3\pi\gamma)$ for states (\ref{Xc}) of the closed
string.

\begin{figure}[bh]
\includegraphics[scale=0.8, trim=29mm 5mm 33mm 7mm]{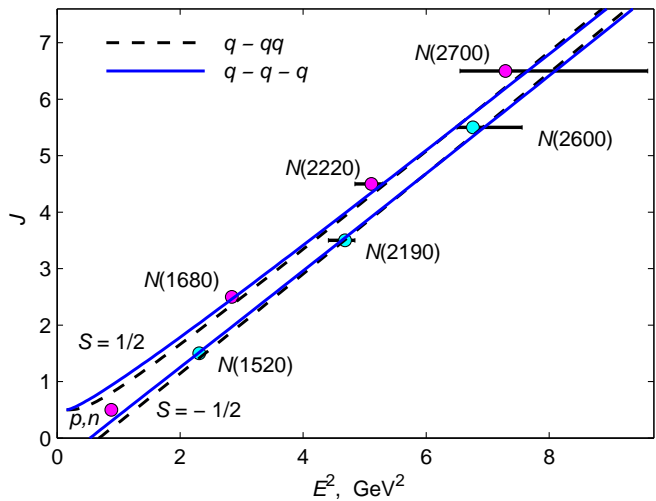}
\caption{Regge trajectories for rotational states (\ref{lrot}) of
the linear baryon model (solid lines) and the quark-diquark model
(dashed lines) }
 \label{RT}
\end{figure}

Fig.~\ref{RT} presents the typical picture of Regge trajectories
for $N$ baryons with $J^P=1/2^+,3/2^-,5/2^+,\dots$ generated by
the linear baryon model (solid lines) in comparison with the
quark-diquark model (dashed lines).

 Here the model parameters are
taken from Ref.~\cite{4B}:
 \be
\gamma=0{.}175\mbox{ GeV}^2,\quad m_q=130\mbox{ MeV},\quad
m_{qq}=2m_q
 \label{gamm12}\ee
 This tension  corresponds to the slope
$\alpha'\simeq0{.}9$ GeV$^{-2}$, effective masses of light quarks
are less than constituent masses \cite{4B,InSh}.

One can see that predictions of the linear baryon model
$q$-$q$-$q$ and the quark-diquark model $q$-$qq$ are rather close
under conditions (\ref{gamm12}). The similar picture takes place
for baryons $\Delta$ and strange baryons \cite{4B,InSh}.

We have shown in Sect.~\ref{Linear} that the rotational states
(\ref{lrot}) of the linear string model are unstable for all
energies on the classic level. But this does not mean
disappearance or terminating corresponding Regge trajectories in
Fig.~\ref{RT}. The straight consequence of this instability is the
contribution to width of a hadron state.

String models describe only excited hadron states with large
orbital momenta $L$. These states are unstable with respect to
strong decays and have rather large width $\Gamma$. In string
interpretation this width is connected with probability of string
breaking; this probability is proportional to the string length
$\ell$ \cite{Kodecay,GuptaR}. The value $\ell$ is proportional to
the string contribution $E_{str}$ to energy $E$ of a hadron state.
For rotational states (\ref{lrot}) and (\ref{Xc}) this
contribution to the expression (\ref{E}) is $E_{str}=q\pi\gamma
a_0$.

Therefore, the component of width $\Gamma_{br}$, connected with
string breaking, is proportional to $E_{str}$ with the factor
$0{.}1$ resulting from particle data \cite{Kodecay,GuptaR,PDG}:
 \be
 \Gamma_{br}\simeq0{.}1\cdot E_{str}=0{.}1\cdot q\pi\gamma  a_0.
 \label{Gambr} \ee

If a state of a string system is unstable with respect to small
disturbances on the classical level, we are to take this
instability into account in the form of additional summand in
width $\Gamma$ of this hadron state:
 \be
 \Gamma=\Gamma_{br}+\Gamma_{inst}.
 \label{Gam} \ee
The contribution  $\Gamma_{inst}$ due to the mentioned instability
is determined by the increment $\xi_2=\xi_2^*$ of exponential
growth (\ref{expps})
 $$
 |\psi|\sim\exp(\xi_2^*\tau)=\exp(\xi_2^*a_0^{-1}t)
 $$
 and relation $t=a_0\tau$ for rotational states (\ref{lrot}) and
 (\ref{Xc}). So for these states
 \be
 \Gamma_{inst}\simeq\frac{\xi_2^*}{a_0}.
 \label{Gaminst} \ee

\begin{figure}[th]
\includegraphics[scale=0.8, trim=1mm 3mm 0mm 0mm]{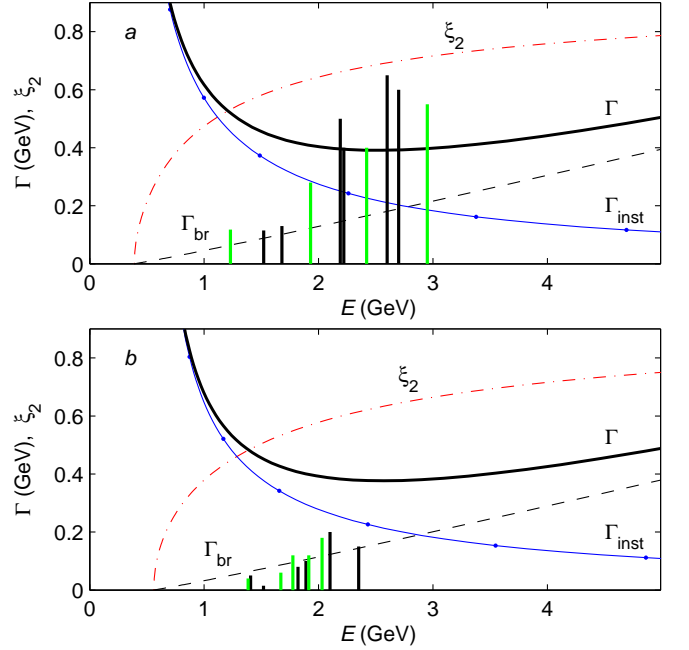}
\caption{Width $\Gamma=\Gamma(E)$ (\ref{Gam}) (solid line), its
summands $\Gamma_{br}$ (\ref{Gambr}) (dashed line) and
$\Gamma_{inst}$ (\ref{Gaminst}) (line with dots) for states
(\ref{lrot}) of the model $q$-$q$-$q$ (a) with parameters
(\ref{gamm12}); (b) with $m_2=300$ MeV}
 \label{Glin}
\end{figure}

The values $\xi_2^*$, $a_0$ and both summands (\ref{Gambr}) and
(\ref{Gaminst}) of the width (\ref{Gam}) depend on energy $E$ of
the hadron state. This dependence for values $\xi_2^*$,
$\Gamma_{inst}$, $\Gamma_{br}$, and $\Gamma$ is calculated for
rotational states (\ref{lrot}) of the model $q$-$q$-$q$,
corresponding to parameters (\ref{gamm12}) for the $N$ baryons in
Fig.~\ref{Glin}.  These graphs are presented in
Fig.~\ref{Glin}{\it a} in comparison with experimental widths of
$N$ and $\Delta$ baryons \cite{PDG} lying on main Regge
trajectories. These widths are shown as the bar graph with dark
bars for $N$ baryons mentioned in Fig.~\ref{RT} and light bars for
baryons $\Delta(1232)$, $\Delta(1930)$, $\Delta(2420)$,
$\Delta(2950)$.

Note that the dimensionless value $\xi_2^*=\xi_2^*(E)$ tends to
zero at $E\to E_{min}=\sum m_j$, but $a_0$ tends to zero more
rapidly, so width $\Gamma_{inst}$ (\ref{Gaminst}) tends to
infinity.

In Fig.~\ref{Glin}{\it b} the similar graphs for strange baryons
with $m_2\equiv m_s=300$ MeV, $m_1=m_3=130$ MeV are presented.
Here dark bars show width of baryons $\Lambda(1405)$,
$\Lambda(1520)$, $\Lambda(1820),\dots$, light bars correspond to
$\Sigma(1385)$, $\Sigma(1670)$, $\Sigma(1775)$, $\Sigma(2030)$.

In the mass range 1 -- 2{.}8 GeV the contribution $\Gamma_{inst}$
(\ref{Gaminst}) due to instability of the linear model  exceeds
$\Gamma_{br}$ and tend to infinity at $E\to E_{min}$. This
behavior contradicts experimental data of baryon's width in the
mentioned mass range: $\Gamma$ tends to zero if $E\to E_{min}$. So
one may conclude that the linear baryon model $q$-$q$-$q$ is not
adequate for describing orbitally excited baryon stated as the
consequence of rotational instability of this model.

\medskip

If we use this approach to the string baryon model Y, we conclude
that instability of rotational states (\ref{roty}) does not change
effective hadron's width $\Gamma$, because linear growth of small
disturbances corresponds to zero contribution $\Gamma_{inst}=0$ in
the increment (factor in the exponent of expression
(\ref{expps})).

Hence for the Y configuration width $\Gamma$ equals $\Gamma_{br}$,
the figure similar to Fig.~\ref{Glin} will have only dashed line
for $\Gamma=\Gamma(E)$ in this case.

\medskip

We mentioned above that unstable central rotational states
(\ref{Xc}) of the closed string, considered in Sect.~\ref{Closed},
may be applied for describing the Pomeron trajectory
 \be
 J\simeq1{.}08+0{.}25E^2
 \label{ReggPom} \ee
 corresponding to possible glueball states
\cite{glY08,MeyerT}.

Estimations of gluon masses on the base of gluon propagator in
lattice calculations \cite{BonnBLW,SilvaO} yield values $m_j$ from
700 to 1000 MeV. We suppose that gluon masses $m_j=750$ MeV and
string tension $\gamma=0{.}175\mbox{ GeV}^2$ corresponds to the
value (\ref{gamm12}). These parameters result in Regge
trajectories $J=J(E^2)$ for states (\ref{Xc}) close to the Pomeron
trajectory (\ref{ReggPom}) \cite{glY08}.

\begin{figure}[t]
\includegraphics[scale=0.8, trim=2mm 3mm 0mm 0mm]{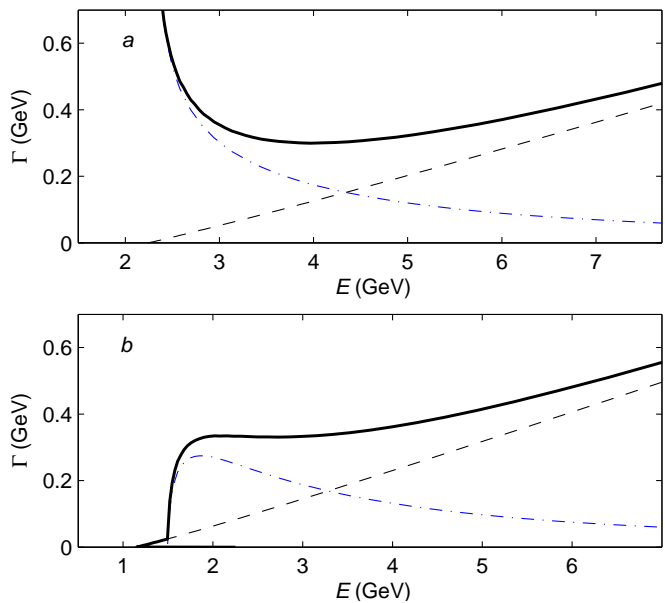}
\caption{Width $\Gamma(E)$ (\ref{Gam}) (solid line) as the sum of
$\Gamma_{br}$ (\ref{Gambr}) (dashed line) and $\Gamma_{inst}$
(\ref{Gaminst}) (dash-dotted line) for central states (\ref{Xc})
of the closed string (a) with parameters $m_j=750$ MeV; (b) with
$m_1=m_2=200$ MeV, $m_3=750$ MeV}
 \label{Gclos}
\end{figure}

In Fig.~\ref{Gclos}{\it a} the total width $\Gamma=\Gamma(E)$
(\ref{Gam}) with its summands $\Gamma_{br}$  and $\Gamma_{inst}$
calculated in Ref.~\cite{Unst09} is presented for central
rotational states (\ref{Xc}). They are unstable for all energies
$E$, if masses $m_j$ are equal. The corresponding width
$\Gamma_{inst}(E)$ tends to infinity in the limit $E\to
E_{min}=\sum m_j$.

Behavior of width $\Gamma(E)$ (\ref{Gam}) for central rotational
states of the system with $m_3>m_1+m_2$ is presented in
Fig.~\ref{Gclos}{\it b}. In this case the threshold effect
(\ref{m3cr}) exists, so the ``instability'' width
$\Gamma_{inst}(E)$ equals zero for energies $E<E_{cr}=2m_3$ (here
it is 1{.}5 GeV). For $E>E_{cr}$  $\Gamma_{inst}(E)$ exceeds
$\Gamma_{br}(E)$ in the certain interval, but if $E$ grows,
$\Gamma_{inst}(E)$ tends to zero and $\Gamma_{br}(E)$ increases.

\section{Conclusion}

The stability problem was solved for classic rotational states
(\ref{roty}) of the Y string baryon model and (\ref{lrot}) of the
linear string baryon model in comparison with states (\ref{Xc})
for the closed string with massive points. It was shown for all
models that the mentioned rotations are unstable, but this
instability has different specific features. For the linear model
spectra of small disturbances for states (\ref{lrot}) contain
complex frequencies for any nonzero values of masses $m_j$. They
are roots of Eqs.~(\ref{lfreq}). These frequencies
$\xi=\xi_1+i\xi_2$ correspond to exponentially growing modes of
disturbances $\psi\sim\exp(\xi_2\tau)$ and, consequently, to
instability of the mentioned rotational states.

The similar behavior takes place for the closed string, but in
this case we have the threshold effect, the central rotational
states (\ref{Xc}) unstable, if the central mass is nonzero and
less than the critical value $m_{cr}$ (\ref{m3cr}). This critical
value equals energy of the string without the central mass.

Rotational instability of the string Y configuration has another
nature. There is no complex frequencies $\xi=\xi_1+i\xi_2$ in the
spectrum (\ref{lfreq}) of small disturbances for states
(\ref{roty}), but this spectrum contains double roots $\xi=\pm\om$
resulting in existence of disturbances (oscillatory modes) with
linearly growing amplitudes.

Instability of classic rotations results in some manifestations in
properties of hadron states, described by the considered string
model. In particular, such a model predicts additional width
$\Gamma_{inst}$ (\ref{Gaminst}) of excited hadrons. Analysis in
Sect.~\ref{Width} shows, that the contribution $\Gamma_{inst}$ in
total width $\Gamma$ (\ref{Gam}), predicted by the linear string
baryon model $q$-$q$-$q$ in the mass range 1 -- 3 GeV is too large
in comparison with experimental data for $N$, $\Delta$ and strange
baryons. These predictions very weakly depend on quark masses
$m_j$ as model parameters. So we conclude, that the linear string
model $q$-$q$-$q$ is unacceptable for describing these baryon
states and we should refuse this model in favor of the
quark-diquark and Y models.

Nevertheless, we can not exclude the $q$-$q$-$q$ configuration as
possible structure of some baryons with anomalously large width or
 a variant of mixing with other configurations.  To make a
definite conclusion for the closed string, considered in
Sect.~\ref{Closed}, we are to have more reliable experimental data
for glueballs and exotic hadrons.

For the Y string baryon model linear (not exponential) growth of
small disturbances corresponds to zero contribution
$\Gamma_{inst}=0$ in the increment of instability  in the exponent
of expression (\ref{expps}). So instability of rotational states
(\ref{roty}) does not give an additional contribution in width
$\Gamma$ of baryons, describing with the Y string model. This
rotational instability of states (\ref{roty}) is not an argument
against application of the Y configuration. But this model has
another drawback, it predicts the slope
$\alpha'=(3\pi\gamma)^{-1}$ for Regge trajectories different from
the value $\alpha'=(2\pi\gamma)^{-1}$ for the string meson model
\cite{Ko,4B}.

\end{document}